\documentclass[11pt]{article}

\usepackage{mathpple}
\usepackage{graphicx}
\usepackage{url}
\usepackage{paralist}
\usepackage{textcomp}

\usepackage{geometry}
\begin{document}

\noindent
\textbf{Preprint of:}\\
A. M. Bra\'{n}czyk, T. A. Nieminen, N. R. Heckenberg
and H. Rubinsztein-Dunlop\\
``Optical trapping of a cube''\\
in R. Sang and J. Dobson (eds),
\textit{Australian Institute of Physics (AIP)
  17th National Congress 2006: Refereed Papers},
Australian Institute of Physics, 2006 (CD-ROM, unpaginated).

\hrulefill

\begin{center}

{\huge
\textbf{Optical trapping of a cube}}

\vspace{2mm}

\textit{A. M. Bra\'{n}czyk, T. A. Nieminen, N. R. Heckenberg
and H. Rubinsztein-Dunlop}\\
School of Physical Sciences, The University of Queensland, Australia

\subsection*{Abstract}

\end{center}

\begin{quote}
The successful development and optimisation of optically-driven
micromachines will be greatly enhanced by the ability to computationally
model the optical forces and torques applied to such devices. In
principle, this can be done by calculating the light-scattering
properties of such devices. However, while fast methods exist for
scattering calculations for spheres and axisymmetric particles,
optically-driven micromachines will almost always be more geometrically
complex.
Fortunately, such micromachines will typically possess a high degree of
symmetry, typically discrete rotational symmetry. Many current designs
for optically-driven micromachines are also mirror-symmetric about a
plane. We show how such symmetries can be used to reduce the
computational time required by orders of magnitude. Similar improvements
are also possible for other highly-symmetric objects such as crystals.
We demonstrate the efficacy of such methods by modelling the optical
trapping of a cube, and show that even simple shapes can function as
optically-driven micromachines.
\end{quote}

\section*{Introduction}

Optical tweezers (Ashkin~\textit{et~al.}~1986) have been deployed
for a variety
of distinct uses: non-contact manipulation of microorganisms,
the measurement of piconewton forces, and as a tool for the study
of a range of microscopic systems, from colloids through to
single molecules. One growing development is the exploitation
of optical torque, which has already seen practical application
(Bishop~\textit{et~al.}~2004; Kn\"{o}ner~\textit{et~al.}~2005).
A major objective, towards which
a number of groups are working, is the development of optically-driven
micromachines (Nieminen~\textit{et~al.}~2006).
A serious impediment, however, is the
difficulty of calculating the expected optical forces and torques
for such micromachines.

The optical forces and torques in optical tweezers arise from
scattering of the trapping beam by the particle. Therefore, the
calculation of these forces and torques is essentially a problem
in computational light scattering. As the particles involved have
dimensions comparable to the wavelength of the light used,
large and small particle approximations, such as geometric optics
and Rayleigh scattering, respectively, are inapplicable. It is
necessary to resort to solution of either the Maxwell equations.
A wide range of methods are available for the solution of such
scattering problems (Kahnert~2003), and, in principle, it should
be possible to use any such method. However, when modelling optical
trapping, one typically wishes to know how the force and torque vary
with position, which required repeated calculations. One basic
question is where in the trap does the particle rest when in
equilibrium---to answer this might require a few dozen calculations
of the force at different positions along the beam axis. To
map the force and torque over a two-dimensional slice through an
optical trap will require approximately a thousand separate calculations
to achieve a reasonable resolution. When one considers the
optical micromanipulation of a complex structure---such as
an optically-driven micromachine---the orientation affects both
the force and the torque, introducing even more degrees of freedom.
Thus, a method that allows rapid repeated calculations is
required for the modelling of optical micromanipulation.

Fortunately, such a method---the \textit{T}-matrix method
(Waterman~1971; Mishchenko~\textit{et~al.}~2004)---is available.
The \textit{T}-matrix
method is more properly a \emph{description} of the scattering properties of
a particle, rather than a method of calculating the scattering properties.
The incident field can be expressed as a set of expansion
coefficients $a_n$ in terms of as 
a sufficiently complete basis set of functions $\psi_n^{(\mathrm{inc})}$,
where $n$ is a mode index labelling the functions, each of which is a
divergence-free solution of the Helmholtz equation:
\begin{equation}
U_\mathrm{inc} = \sum_n^\infty a_n \psi_n^{(\mathrm{inc})}.
\label{exp1}
\end{equation}
Similarly, we can write the scattered wave, in terms of a basis set:
$\psi_k^{(\mathrm{scat})}$,
\begin{equation}
U_\mathrm{scat} = \sum_k^\infty p_k \psi_k^{(\mathrm{scat})},
\label{exp2}
\end{equation}
where $p_k$ are the expansion coefficients for the scattered wave.
As long as the electromagnetic or optical properties of the scatterer
are linear, the relationship between the two can be written as a
simple matrix equation
\begin{equation}
p_k = \sum_n^\infty T_{kn} a_n
\end{equation}
or,  in more concise notation,
\begin{equation}
\mathbf{P} = \mathbf{T} \mathbf{A}
\end{equation}
where $T_{kn}$ are the elements of the \textit{T}-matrix. Thus, the
\textit{T}-matrix formalism is a Hilbert basis description of
the scattering properties of the particle, with the
\textit{T}-matrix depending only on the properties of the
particle---its composition,  size,
shape, and orientation---and the wavelength, and is otherwise
independent of the incident field. As a result,
the \textit{T}-matrix only needs to be calculated once for a particular
particle, after which it can be used for rapid repeated calculations
of the optical force and torque (Nieminen~\textit{et~al.}~2004b).

In the simplest case, that of a homogeneous isotropic sphere, the
\textit{T}-matrix is given by the analytical Lorenz--Mie
solution (Lorenz~1890; Mie~1908), while more complex cases
require computational solution. For homogeneous isotropic
axisymmetric particles of simple shape, such as spheroids or cylinders,
this can be done very rapidly using the extended boundary
condition method, also known as the null-field method
(Tsang~\textit{et~al.}~2001),
since surface integrals over the particle reduce to one dimension.
For more complex shapes, the computational time required increases
greatly. However, symmetries such as discrete rotational symmetry
or mirror symmetry can be used to reduce the time required
(Kahnert~2005). Notably, these are exactly the symmetries typical
of most optically-driven micromachine designs. We will proceed to
use a cube as an example of such optimisation, and in the process
show that even simple shapes can function as optically-driven micromachines.

\section*{Exploiting the Symmetry of a Cube}

For a compact scatterer, the \textit{T}-matrix method is
usually implemented with vector spherical wavefunctions (VSWFs) as
the basis functions, which fall into two groups, TE and TM.
These are usually written as $\mathbf{M}_{nm}$ and $\mathbf{N}_{nm}$,
respectively, where $n$ is the radial mode index and $m$ is the azimuthal
mode index. The properties that are of importance when optimising the
calculation of a \textit{T}-matrix by exploiting the symmetry properties
of a scatterer are the parity and rotational symmetries.

We employ a point-matching method that allows us to calculate
the \textit{T}-matrix column-by-column considering only a single
incident VSWF at a time (Nieminen~\textit{et~al.}~2004b). Each column
requires the solution of an overdetermined linear system, and the
time required by this is the dominant component in the overall computational
time. Solution of linear systems typically scales as $N^3$, where $N$ is
the number of unknowns.

\subsection*{Parity}

Each individual VSWF has either odd or even parity with respect to the
$xy$-plane. That is, the magnitude of the electric field is symmetric
relative to this plane, and the phase is either the same or differs
by $\pi$, such that $\mathbf{E}(x,y,z) = \mathbf{E}(x,y,-z)$ (even
parity) or $\mathbf{E}(x,y,z) = -\mathbf{E}(x,y,-z)$ (odd parity).
TE VSWFs have odd parity when $n+m$ is odd, and even otherwise. TM
VSWFs have odd parity when $n+m$ is even, and even otherwise.
When the scatterer is mirror symmetric about the $xy$-plane, the
parity of the incident field is unchanged on scattering, and thus
the scattered field consists only of modes of the same parity as
the incident field. Accordingly, only half of the total number of
scattered field modes need to be included in the linear system,
halving the number of unknowns, $N$, with a corresponding reduction
in computational time.

\subsection*{Rotational symmetry}

Each individual VSWF has an azimuthal dependence of
$\exp(\mathrm{i}m\phi)$. If a scatterer possesses discrete rotational
symmetry of order $p$, this effectively provides a periodic boundary
condition, determining the periodicity with respect to the
azimuthal angle $\phi$ that the scattered field can possess. From
Floquet's theorem, the allowed azimuthal mode indices for the
scattered field are
\begin{equation}
m_\mathrm{scat} = m_\mathrm{inc} + ip
\end{equation}
where $i$ is an integer. This is analagous to the generation of
a discrete spectrum of scattered plane waves by a grating.
If the particle has no rotational symmetry (ie $p=1$), then
coupling to all azimuthal modes occurs. For an axisymmetric
particle, $p=\infty$ and $m_\mathrm{scat} = m_\mathrm{inc}$,
which is widely used when calculating scattering by such particles.
For a cube, $p=4$, and $m_\mathrm{scat} = m_\mathrm{inc},
m_\mathrm{inc} \pm 4, m_\mathrm{inc} \pm 8, ...$.
As a result, the number of unknown in the linear system is reduced
to approximately $1/4$.

Make use of both symmetries together reduces the number of unknowns
by a factor of eight, with a considerable savings in computational time.
For example, a cube with faces two wavelengths across required 30
minutes for the calculation of the \textit{T}-matrix on a 32 bit
single-processor 3GHz PC, as compared with 30 hours without the
symmetry optimisations.

\section*{Optical Trapping of a Cube}

\begin{figure}[htb]
\begin{center}

\includegraphics[width=0.6\columnwidth]{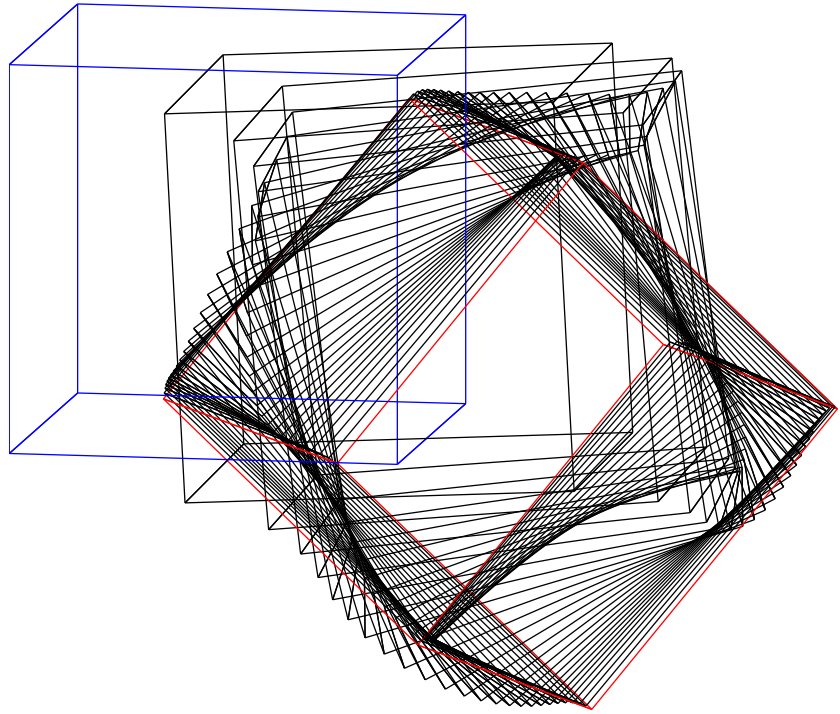}

\caption{Optical trapping of a cube}

\end{center}
\end{figure}

Figure 1 shows the optical trapping of a cube. The initial position
of the cube is shown by the blue frame, and the equilibrium position
by the red frame. At each position, the optical force and torque are
calculated, and the motion of the cube found. The cube is assumed to
be moving at terminal speed at all times (the time constant for
approach to terminal speed in optical tweezers is typically
on the order of 0.1\,{\textmu}m (Nieminen~\textit{et~al.}~2001),
which is much smaller than the time steps used to calculate the
motion of the cube).

As the discrete rotational symmetry of a cube is typical of the
rotational symmetry of optical driven micromachines
(Nieminen~\textit{et~al.}~2006), and the optical torque is
determined by the rotational symmetry of a scatterer
(Nieminen~\textit{et~al.}~2004a), a cube can be expected to
rotate when illuminated by a beam carrying angular momentum.
To test this, the cube is place initially on the beam axis,
at the focus of the beam. The beam is circularly polarised, and therefore
carries spin angular momentum of $\hbar$ per photon.
As can be seen in figure 2, the cube is rapidly pushed into the equilibrium
position along the beam axis, where it spins due to the transfer of
angular momentum from the beam to the cube. Although the face-up position
is an unstable equilibrium, any torque acting to bring the cube into the
stable corner-up position seen in figure 1 is too small to have any
visible effect over the duration of the simulation. The cube is shown
both from the side and from below.

\begin{figure}[htb]
\begin{center}
~
\hfill
\includegraphics[width=0.3\columnwidth]{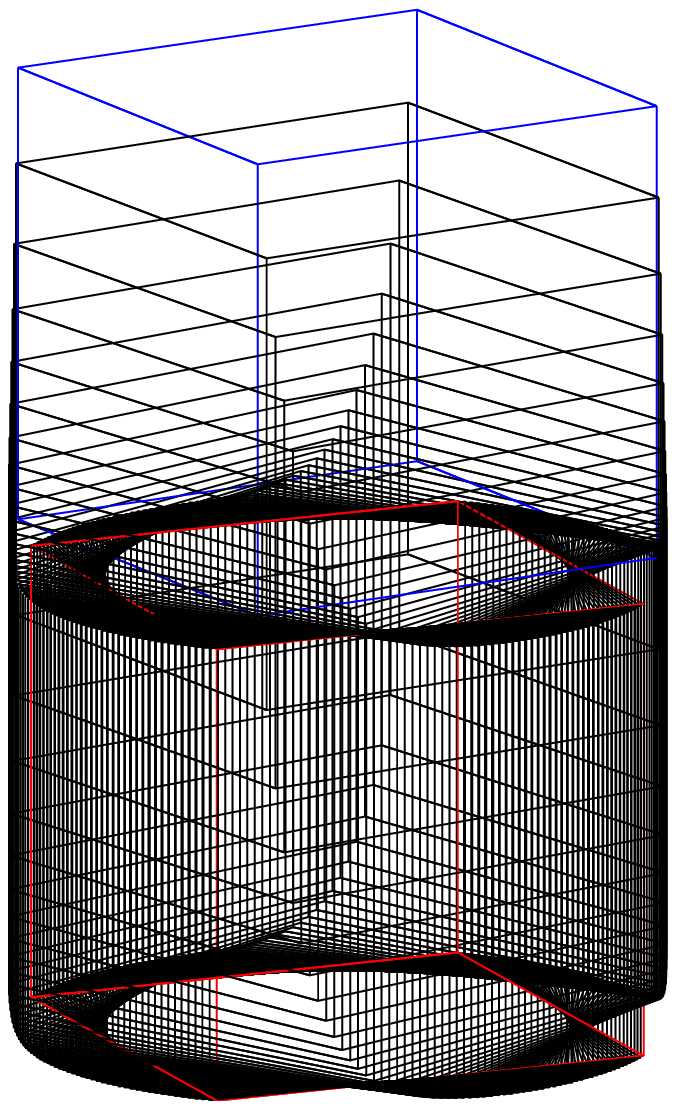}
\hfill
\includegraphics[width=0.3\columnwidth]{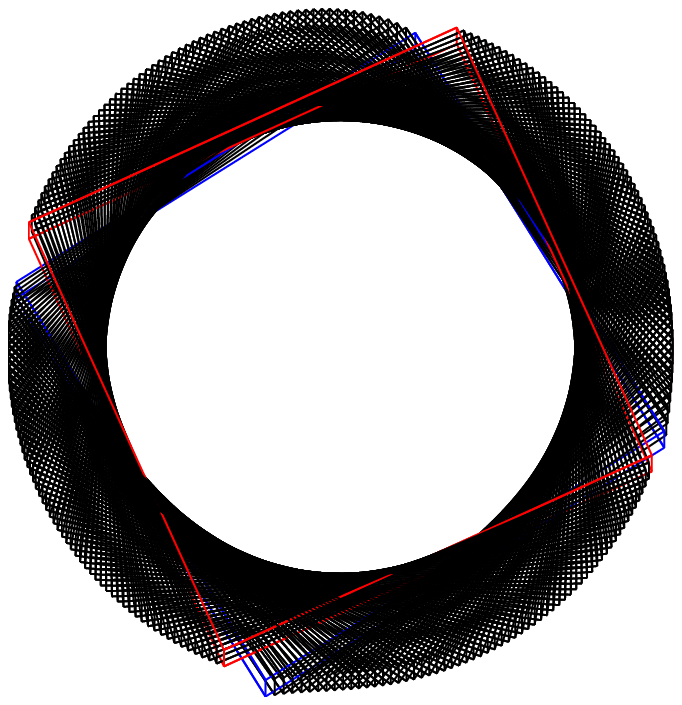}
\hfill
~

\caption{Rotation of a cube}
\end{center}
\end{figure}

\section*{Refererences}

\setlength{\parindent}{0pt}

\hangindent=1cm
Ashkin, A., Dziedzic, J.M., Bjorkholm, J.E. and Chu, S. (1986).
Observation  of  a  single-beam  gradient  force optical trap for
dielectric particles.
\textit{Optics Letters} \textbf{11}, 288-90.

\hangindent=1cm
Bishop, A.I., Nieminen, T.A., Heckenberg, N.R. and
Rubinsztein-Dunlop, H. (2004).
Optical microrheology using rotating laser-trapped particles.
\textit{Physical Review Letters} \textbf{92}, 198104.

\hangindent=1cm
Kahnert, F.M. (2003).
Numerical methods in electromagnetic scattering theory.
\textit{Journal of Quantitative Spectroscopy and Radiative Transfer}
\textbf{79-80}, 775-824.

\hangindent=1cm
Kahnert, M. (2005).
Irreducible  representations  of  finite  groups  in the \textit{T}-matrix
formulation of the electromagnetic scattering problem.
\textit{Journal  of  the  Optical  Society  of  America A}
\textbf{22}, 1187-99.

\hangindent=1cm
Kn\"{o}ner, G., Parkin, S., Heckenberg, N.R. and
Rubinsztein-Dunlop, H. (2005).
Characterization  of  optically  driven  fluid stress fields with
optical tweezers.
\textit{Physical Review E} \textbf{72}, 031507.

\hangindent=1cm
Lorenz, L. (1890).
Lysbev{\ae}gelsen i og uden for en af plane Lysb{\o}lger belyst Kugle.
\textit{Videnskabernes Selskabs Skrifter} \textbf{6}, 2-62.

\hangindent=1cm
Mie, G. (1908).
Beitr\"{a}ge   zur   Optik   tr\"{u}ber   Medien,   speziell  kolloidaler
Metall\"{o}sungen. \textit{Annalen der Physik} \textbf{25}, 377-445.

\hangindent=1cm
Mishchenko, M.I., Videen, G., Babenko, V.A., Khlebtsov, N.G. and Wriedt, T.
(2004).
\textit{T}-matrix  theory  of  electromagnetic scattering by particles and
its applications: a comprehensive reference database.
\textit{Journal  of  Quantitative  Spectroscopy and Radiative Transfer}
\textbf{88}, 357-406.

\hangindent=1cm
Nieminen, T.A., Rubinsztein-Dunlop, H., Heckenberg, N.R. and Bishop, A.I.
(2001).  Numerical modelling of optical trapping.
\textit{Computer Physics Communications} \textbf{142}, 468-71.

\hangindent=1cm
Nieminen, T.A., Rubinsztein Dunlop, H. and Heckenberg, N.R. (2003).
Calculation   of   the  \textit{T}-matrix:   general  considerations  and
application of the point-matching method.
\textit{Journal of Quantitative Spectroscopy and Radiative Transfer}
\textbf{79-80}, 1019-29.

\hangindent=1cm
Nieminen, T.A., Parkin, S.J., Heckenberg, N.R. and
Rubinsztein-Dunlop, H. (2004a).
Optical torque and symmetry.
\textit{Proceedings of SPIE} \textbf{5514}, 254-63.

\hangindent=1cm
Nieminen, T.A., Heckenberg, N.R. and Rubinsztein-Dunlop, H. (2004b).
Computational modelling of optical tweezers.
\textit{Proceedings of SPIE} \textbf{5514}, 514-23.

\hangindent=1cm
Nieminen, T.A., Higuet, J., Kn\"{o}ner, G., Loke, V.L.Y., Parkin, S.,
Singer, W., Heckenberg, N.R. and Rubinsztein-Dunlop, H. (2006).
Optically driven micromachines: progress and prospects.
\textit{Proceedings of SPIE} \textbf{6038}, 237-45.

\hangindent=1cm
Tsang, L., Kong, J.A., Ding, K.H. and Ao, C.O. (2001).
Scattering of Electromagnetic Waves: Numerical Simulation.
Wiley, New York.

\hangindent=1cm
Waterman, P.C. (1971).
Symmetry, unitarity, and geometry in electromagnetic scattering.
\textit{Physical Review D} \textbf{3}, 825-39.

\end{document}